\PassOptionsToPackage{table,xcdraw}{xcolor}
\documentclass[11pt,a4paper]{article}
\pdfoutput=1
\usepackage{jheppub}
\usepackage{gensymb}
\DeclareSymbolFont{matha}{OML}{txmi}{m}{it}
\DeclareMathSymbol{\varv}{\mathord}{matha}{118}
\usepackage{amsmath}
\usepackage{amssymb}
\usepackage{mathrsfs}
\usepackage{graphicx}
\usepackage{subfigure}
\usepackage{pstricks}
\usepackage{bm}
\usepackage{pbox}
\usepackage{placeins}
\usepackage{booktabs}
\usepackage[T1]{fontenc}
\usepackage{footnote}
\usepackage{pdfpages}
\usepackage{hhline}
\usepackage{multirow}
\usepackage{multicol}
\usepackage{enumitem}
\usepackage[toc,page]{appendix}
\allowdisplaybreaks
\usepackage{comment}
\usepackage{float}
\usepackage{slashed}
\usepackage{xcolor}
\usepackage{textcomp}
\usepackage{gensymb}
\usepackage{multirow}
\usepackage[numbers]{natbib}
\usepackage{notoccite}
\usepackage{dsfont}
\usepackage{rotating}
\usepackage{tabularx}
\usepackage{soul}
\usepackage{tabu}
\usepackage{array}
\usepackage{cancel}
\usepackage{longtable}

\FloatBarrier

\usepackage[many]{tcolorbox}

\renewcommand{\thetable}{\Roman{table}}

\renewcommand{\thetable}{\Roman{table}}
 
\newcommand{\vev}[1]{\langle#1\rangle}
\newcommand{\GeV}{\mathrm{GeV}}
\newcommand{\tr}{\mathrm{Tr}}
\newcommand{\ee}[1]{\times 10^{#1}}
\newcommand{\diag}{\mathrm {diag}}
\newcommand{\diff}{\mathrm d}

\definecolor{MyDarkBlue}{rgb}{0.1, 0.1, 0.8} 
\definecolor{MyLightBlue}{rgb}{0.22,0.51,0.9}
\definecolor{MyGreen}{rgb}{0.0, 0.5, 0.0}
\definecolor{BrickRed}{rgb}{0.8, 0.25, 0.33}
\definecolor{Aqua}{rgb}{0.1, 0.6, 0.8}
\definecolor{Olive}{rgb}{ 	0, .42, .3}

\RequirePackage{hyperref}
\hypersetup{colorlinks, citecolor=BrickRed,linkcolor=blue, urlcolor=MyLightBlue}

\makeatletter
\gdef\@fpheader{}
\makeatother
\begin{document}

\title{\bf A Generalised Missing Partner Mechanism \\for \texorpdfstring{$SU(5)$}{SU(5)} GUT Inflation}

\author{Stefan Antusch,}
\author{Kevin Hinze,}
\author{Shaikh Saad,}
\author{and Jonathan Steiner}

\affiliation{Department of Physics, University of Basel,\\ Klingelbergstrasse\ 82, CH-4056 Basel, Switzerland}

\emailAdd{stefan.antusch@unibas.ch, kevin.hinze@unibas.ch, shaikh.saad@unibas.ch, jonathan.steiner@stud.unibas.ch}
\abstract{We generalise the Missing Partner Mechanism to split the electron-like states from the coloured ones of vectorlike $SU(5)$ 10-plets without fine-tuning. Together with the extra light weak doublets from the Double Missing Partner Mechanism (DMPM), this realises gauge coupling unification in the presence of a light weak triplet and colour octet, the characteristic light relics from the adjoint in $SU(5)$ GUT Inflation models. Additionally, we show how the vectorlike 10-plets may generate realistic fermion masses while the DMPM ensures that dimension five nucleon decay is suppressed. A discovery of the light relic states at future colliders would provide a ``smoking gun'' signal of the scenario.
}
\maketitle

\section{Introduction}

Grand Unified Theories (GUTs)~\cite{1973PhRvL..31..661P,1974PhRvD..10..275P, 1974PhRvL..32..438G,1974PhRvL..33..451G,CALT-68-467} are among the most attractive candidates for physics beyond the SM (BSM). On the other hand, cosmic inflation~\cite{STAROBINSKY198099,1981PhRvD..23..347G,1982PhLB..108..389L,1982PhRvL..48.1220A} -- the accelerated expansion of the universe at early times -- can explain why the universe appears to be flat and homogeneous on large scales. Interestingly, the three fundamental forces of the SM tend to unify at a high scale, which is similar to the energy scale where cosmic inflation may take place. This remarkable coincidence indicates that  inflation could be linked to GUTs. Furthermore, inflation provides a natural mechanism to dilute away the unwanted monopoles created abundantly during GUT symmetry breaking~\cite{1981PhRvD..23..347G,1982PhLB..108..389L}. 

In this work, we focus on supersymmetric (SUSY) $SU(5)$ GUTs. To embed inflation, one may distinguish three characteristic approaches where different types of GUT representations mainly play the role of the inflaton. 
One of them is supersymmetric ``Hybrid Inflation''~\cite{Linde:1991km,astro-ph/9307002,hep-ph/9406319,astro-ph/9401011,1104.4143}, where the inflaton is a gauge singlet superfield. It turns out that in the minimal version, commonly termed as ``Standard SUSY Hybrid Inflation'', the monopole problem within $SU(5)$ GUTs is not resolved. Two variations of it, namely, ``Shifted Hybrid Inflation''~\cite{hep-ph/0002151} and ``Smooth Hybrid Inflation''~\cite{hep-ph/9506325,1202.0011} overcome this shortcoming. In a second class of models -- ``Tribrid Inflation''~\cite{hep-ph/0411298,0808.2425,0908.1694,0902.2934,1003.3233,1007.0708,1102.0093} -- a D-flat combination of  matter fields -- typically gauge non-singlets -- plays the role of the inflaton. A slight modification of the original proposal is needed to build realistic models that do not suffer from the monopole problem, which is known as  ``Pseudosmooth Tribrid Inflation''~\cite{1205.0809}. A third class of models, known as  ``New Inflation''~\cite{hep-ph/0403294}, utilises the GUT breaking field as the inflaton. Here, inflation takes place while the GUT breaking field is rolling slowly in the flat region of a hilltop-shaped potential where the GUT symmetry is already broken, diluting away the monopoles. 

In all these classes of models, the required form of the superpotential for successful inflation to take place is typically guaranteed by a global $R$-symmetry, $U(1)_R$, and depending on the scenario by an additional $\mathbb{Z}_N$ symmetry. An immediate consequence of this $R$-symmetric $SU(5)$ GUTs is that weak-triplet and colour-octet fields from the adjoint Higgs remain light~\cite{hep-ph/0511097,1109.4797}. This spoils the successful gauge coupling unification (GCU) of the minimal supersymmetric Standard Model (MSSM). One known solution to this problem is that if some components, namely, weak-doublet and singly-charged fields, of additional vectorlike families, reside close to the SUSY breaking scale, gauge coupling unification can be restored~\cite{1010.3657}. However, their presence may seem ad-hoc and, moreover, without explanation this splitting of GUT multiplets would lead to undesirable fine-turning. 
 
Another challenge of SUSY GUTs is that, on top of the usual (dimension 6) gauge-mediated nucleon decay, one gets potentially much more dangerous (dimension 5) nucleon decay modes by exchanging the heavy colour-triplet Higgs~\cite{Dimopoulos:1981zb,Sakai:1981pk}. Therefore, these colour-triplets need to be superheavy, whereas their partners, the weak-doublets, must reside at the electroweak scale. This so-called ``Doublet-Triplet Splitting'' (DTS) problem~\cite{Randall:1995sh,Yamashita:2011an} can introduce another potential source of severe fine-tuning in the theory.

In this work we propose a way to embed inflation into $SU(5)$ GUTs while resolving the above mentioned (and further) challenges of GUT Inflation scenarios without excessive fine-tuning. To this end, we extend the   ``Missing Partner Mechanism'' (MPM)~\cite{1982PhLB..115..380M,HUTP-82/A014}, or more specifically its improved version, the ``Double Missing Partner Mechanism'' (DMPM)~\cite{hep-ph/9406417,hep-ph/9611277,1405.6962}, which are attractive ways of resolving the DTS problem and suppressing dimension 5 nucleon decay, to a ``Generalised Missing Partner Mechanism'' (GMPM). In addition to naturally splitting the Higgs doublets from the triplets, it also provides light weak-doublet and singly-charged fields from vectorlike states that reinstate gauge coupling unification. Intriguingly, these vectorlike chiral supermultiplets, along with the fields employed to realise the GMPM, simultaneously correct the wrong mass relations between the down-type quark and charged-lepton sectors. The light relic states provide a characteristic feature of the scenario, which allows to potentially test it at future colliders.

This paper is organised in the following way. In Sec.~\ref{sec:inf}, we briefly review inflation in $SU(5)$ GUTs, in Sec.~\ref{:sec:TandO} the origin of massless relics is discussed. In Sec.~\ref{sec:GCU} we present our solution to gauge coupling unification and we show how realistic fermion masses can arise in Sec.~\ref{sec:FM}.  Phenomenological aspects of the light relics along with relevant collider bounds are summarised in Sec.~\ref{sec:pheno}, and finally we conclude in Sec.~\ref{sec:con}.

\section{Inflation in \texorpdfstring{$SU(5)$}{SU(5)} GUTs}\label{sec:inf}
This section briefly summarises three classes of realistic inflationary models that are consistent within SUSY $SU(5)$ GUTs. In all these models, the superpotential relevant for the inflationary dynamics takes a simple form,
\begin{align}
W_\mathrm{inflation}=\varkappa S\left( f_1(\Phi)-M^2 \right) + f_2(\Phi)f_3(N_i), \label{eq:W-inf}     
\end{align}
where $S$ is a GUT singlet superfield, $\Phi$  is an adjoint superfield that breaks the GUT symmetry, $N_i$ are matter superfields that are potentially also GUT non-singlets. 
$\varkappa$ is a dimensionless coupling constant and $M$ is a parameter with mass dimension one. 
For consistency, the holomorphic functions $f_k$ must be such that the resulting field combinations are GUT singlets. This form of the superpotential, Eq.~\eqref{eq:W-inf}, is guaranteed by $R$-symmetry, under which we demand that  
\begin{align}
W\to e^{i2\theta}W, \;\;\;
S\to e^{i2\theta}S, \;\;\;
f_{1,2}(\Phi)\to f_{1,2}(\Phi), \;\;\;
f_3(N_i)\to e^{i2\theta}f_3(N_i).
\end{align}
The explicit expressions of $f_i$ depends on the model class and the specific realisation of inflation, to be discussed below. 

\vspace{0.5cm}
\textbf{1.\;GUT-Singlet Field as the Inflaton:}--
In this class of models, the gauge singlet field, $S$, mainly plays the role of the inflaton. In the minimal/standard version of Hybrid Inflation, the functions $f_i$ are taken to be $f_1=\Phi^2$ and $f_{2,3}=0$,  
leading to $W_\mathrm{hybrid}=\varkappa S(\Phi^2-M^2)$ \cite{Dvali:1994ms,Copeland:1994vg,Linde:1997sj}.\footnote{Here and in the following we use $\Phi^n$ as a shorthand notation for the all possible terms that contract the indices of the adjoint to form a singlet, here $\Phi^2 \equiv \tr (\Phi^2)$.} 
In the very early universe, with a large vacuum expectation value (vev) 
of the $S$ field beyond a critical value, $\Phi$ is stabilised at zero. Consequently, the GUT symmetry remains unbroken, and within global SUSY the potential is flat at the tree level along the $S$ direction. The energy density is then dominated by the false vacuum energy density, $V\sim \varkappa^2 M^4$, leading to an exponentially expanding universe. After including radiative corrections, corrections from SUSY breaking and K\"ahler potential terms, a small tilt is generated such that the $S$ field rolls slowly towards the above-mentioned critical value, giving rise to a successful inflationary scenario \cite{BasteroGil:2006cm,Rehman:2009nq,Nakayama:2010xf,Antusch:2012jc,Buchmuller:2014epa,Schmitz:2018nhb}.\footnote{Also for the other scenarios to be discussed below, we will always assume that corrections from SUSY breaking and K\"ahler potential terms can be adjusted such that the inflationary predictions are consistent with CMB observations.} When $S$ reaches the critical value, it destabilises $\Phi$, which then  quickly rolls to its minimum with non-zero vev, breaking the GUT symmetry and ending inflation. Since the $SU(5)$ GUT symmetry breaking takes place at the end of inflation, monopoles are produced. While this rules out the standard version of Hybrid Inflation in $SU(5)$, variants of it can be viable, as we now discuss:

\begin{enumerate}
\item[$\bullet$] \textbf{Shifted Hybrid Inflation:} In Shifted Hybrid Inflation~\cite{hep-ph/0002151,1010.3657}, non-renormalizable terms are added to the superpotential, introducing a non-trivial flat direction viable for inflation, along which the GUT symmetry is already spontaneously broken. Monopoles may have been produced before inflation, but are ``inflated away'', resolving the monopole problem. 
The superpotential in Shifted Hybrid Inflation takes e.g.\ the form
\begin{equation}
W_\mathrm{shifted-hybrid}=\varkappa_1 \, S(\Phi^2-M^2)
- \underbrace{ \varkappa_2 \, \frac{S\Phi^{2+n}}{\Lambda^n} }_{\equiv \delta W}, \label{eq:ShiftedHyInf}
\end{equation}
where, compared to standard Hybrid Inflation, the only new term added is $\delta W$. The leading order non-renormalizable term corresponds to $n=1$.  As before, this specific form of the superpotential is guaranteed by the $R$-symmetry. $\Lambda$ denotes the cutoff scale for the respective effective operator.
The presence of the non-renormalizable  term(s) in Eq.~\eqref{eq:ShiftedHyInf} ensures that for constant $|S|$, in addition to the $\Phi=0$ minimum, a local minimum with $\Phi\neq 0$ is also possible.  
Also for this non-trivial minimum, the potential is tree-level flat (apart from K\"ahler corrections and corrections from the SUSY breaking sector) in the $S$ direction, 
providing a suitable trajectory for inflation. Since the GUT symmetry is already spontaneously broken during inflation, monopoles that may have formed earlier are efficiently ``inflated away'', i.e.\  diluted by inflation, thus resolving the monopole problem.

\item[$\bullet$] \textbf{Smooth Hybrid Inflation:}  The strategy to avoid the monopole problem for Smooth Hybrid Inflation~\cite{hep-ph/9506325,1202.0011} is similar to the shifted case, but the realisation is somewhat different. In both scenarios, inflation occurs along a shifted track where the GUT symmetry is already broken during inflation, resolving the monopole problem. The difference is that in the smooth superpotential, the renormalizable trilinear coupling, $S\Phi^2$, is forbidden by the imposition of a   $\mathbb{Z}_m$ symmetry. Under this discrete symmetry, $\Phi^2$ transforms non-trivially, whereas $\Phi^m$ is invariant (we define, $m=2+n$, with $n>0$). Therefore, the superpotential takes the form
\begin{equation}
    W_\mathrm{smooth-hybrid}=\varkappa \, S\left(\frac{\Phi^{2+n}}{\Lambda^n}-M^2\right).\label{smooth}
\end{equation}   
For any fixed value of $S$, the potential (in the $\Phi$ direction) has a local maximum at $\Phi=0$, and an absolute minimum for $\Phi\neq 0$. This implies that inflation ends smoothly without any ``waterfall'' where the $\Phi$ direction would become tachyonic. Another difference compared to the shifted case is that the inflationary trajectory already has a slope at tree-level.   

\end{enumerate}

\textbf{2.\;Matter Field as the Inflaton:}-- 
In this class of models, known as the Tribrid Inflation~\cite{hep-ph/0411298,0808.2425,0908.1694,1003.3233,1910.07554}, a D-flat direction of matter fields that are potentially gauge non-singlets plays the role of the inflaton.  In the first model of
Sneutrino Tribrid Inflation, proposed in Ref.~\cite{hep-ph/0411298}, the symmetry gets broken at the end of inflation. Hence, when applied to an $SU(5)$ GUT, the monopole problem is not solved. In this framework, the $S$ field stays at zero during inflation. It can be neglected for the inflationary dynamics, but it contributes to the large vacuum energy by its F-term. Stabilising $S$ at zero is guaranteed by generating a large mass of order Hubble scale via non-minimal terms in the K\"{a}hler potential. When the matter field $N$ has a vev above a critical value, it stabilises $\Phi$ at zero. The $N$ field direction is tree-level flat, and suitable for realising inflation.  
When $\Phi$ reaches the critical value, a ``waterfall'' takes place that ends inflation and the GUT symmetry gets broken. For $SU(5)$ GUT Inflation, one needs to choose a variant that solves the monopole problem:

\begin{enumerate}
\item[$\bullet$] \textbf{Pseudosmooth Tribrid Inflation:} 
In Pseudosmooth Tribrid inflation \cite{1205.0809,Masoud:2019gxx}, the superpotential has the following form:
\begin{equation}
W_\mathrm{pseudosmooth-tribrid}=\varkappa_1 \,S\left(\frac{\Phi^{2+n}}{\Lambda^n} -M^2\right)-\varkappa_2  \:\frac{\Phi^{1+p}}{\Lambda^{p+q}}N^{2+q}. \label{eq:tribrid}
\end{equation}
In addition to the $U(1)_R$ symmetry, the form of this superpotential is ensured by introducing a $\mathbb{Z}_m$ symmetry such that both $\Phi$ and $N$ carry charge 1 under it, with $n=m-2, \, p+q=m-3$. In this mechanism, for inflation to happen, a ``shifted smooth'' track is employed with the GUT symmetry already broken during inflation, hence monopoles produced earlier are inflated away.  
Though inflation occurs along an initially smooth trajectory, unlike in Smooth Hybrid Inflation there is a ``waterfall'' transition to a unique minimum where $N = 0$. Because of this the mechanism is named pseudosmooth. 
\end{enumerate}

\textbf{3.\;GUT-breaking Field as the Inflaton:}-- In this class of models, also known as ``New Inflation'', the scalar component of the GUT Higgs superfield breaking $SU(5)$ by its vev plays the role of the inflaton. 
In this setup~\cite{hep-ph/0403294,Antusch:2013eca}, the superpotential takes the same form as the Smooth Hybrid Inflation model, i.e.,
\begin{equation}
    W_\mathrm{new-inflation}=\varkappa \, S\left(\frac{\Phi^{2+n}}{\Lambda^{n}}-M^2\right),\label{newINF}
\end{equation}
with $n\geq 2$.  However, the primary difference is that in Smooth Hybrid Inflation, the singlet field $S$ mainly plays the role of the inflaton, whereas in New Inflation scenario, the GUT breaking adjoint field itself is the inflaton. The structure of Eq.~\eqref{newINF} is again guaranteed by $U(1)_R\times \mathbb{Z}_m$ symmetry (where $m=n+2$). $S$ is again stabilised at zero via non-minimal terms in the K\"{a}hler potential. The scalar potential obtained from the superpotential Eq.~\eqref{newINF} features a flat hilltop as in the New Inflation models \cite{hep-ph/0403294,1804.07619}. 
The monopole problem is resolved since the GUT symmetry is already broken by the vev of $\Phi$ while inflation is ongoing.

We like to emphasise that in all the $SU(5)$ GUT Inflation scenarios outlined above, an $R$-symmetry plays a crucial role in guaranteeing the form of the inflaton potential, with a large vacuum energy and simultaneously a sufficiently flat trajectory for inflation. This $R$-symmetry, usually realised as a $U(1)_R$ or potentially by a discrete subgroup of it, leads to light components of the adjoint $\Phi$, as we now discuss.

\section{Masses of the light triplets and octets}\label{:sec:TandO}

As emphasised above, within the supersymmetric framework, the $R$-symmetry plays an essential role in realizing inflation. However, a consequence of this symmetry is that it leads, in the limit of unbroken global SUSY, to massless states originating from the adjoint superfield, which breaks the GUT symmetry. Recall that  under the SM, the adjoint of $SU(5)$ decomposes in the  following way:
\begin{equation}
    \Phi(24)=\mathscr O(8,1)_0\oplus\mathscr T(1,3)_0\oplus X(3,2)_{-5/6}\oplus\overline X(\overline 3,2)_{5/6}\oplus\phi(1,1)_0.\label{eq:branching24}
\end{equation}
Once $SU(5)$ is broken, the $X+\overline{X}$ supermultiplets get eaten up. The requirement that the scalar potential is minimised guarantees a GUT scale mass for the SM singlet, and due to SUSY, also for its fermionic partner. However, both the scalar and the fermionic  partners of the colour octet, $\mathscr O$, and the weak triplet, $\mathscr T$, remain massless. This can be easily seen from the inflationary superpotential given in  Eq.~\eqref{eq:W-inf}. Let us consider the corresponding fermionic part of the Lagrangian for the octet and the triplet,
\begin{multline}
    \mathcal L\supset S\Big[\tilde f_{1,\mathscr O}(\vev{\Phi})\tilde {\mathscr 
 O}\tilde{\mathscr  O}+\tilde f_{1,\mathscr T} (\vev{\Phi}) \tilde{\mathscr  T}\tilde{\mathscr  T}\Big]+f_3(N_i) \Big[\tilde f_{2,\mathscr O}(\vev{\Phi})\tilde{\mathscr  O}\tilde{\mathscr  O}+\tilde f_{2,\mathscr T} (\vev{\Phi}) \tilde{\mathscr  T}\tilde{\mathscr  T}\Big]\\+H.c. \;.\label{eq:fermionmass}
\end{multline}
In the above, $\tilde{\mathscr  O}$ and $\tilde{\mathscr  T}$ stand for the fermionic components of the octet and triplet while $-2\tilde f_{i,\digamma}\equiv\partial^2_\digamma f_i$. In the global SUSY minimum, we have $\vev S=\vev{N_i}=0$, resulting (for cases where $f_3$ goes to zero for $N_i = 0$) in massless fermionic as well as scalar states. Note that this is a quite general feature of $U(1)_R$ symmetric GUTs, as shown in Refs.~\cite{hep-ph/0511097,1109.4797}.

Supersymmetry, however, must be broken in nature. Subsequently, the above-mentioned massless states obtain masses of the order of the gravitino mass, i.e.,  $m_{\tilde{\mathscr  O},\tilde{\mathscr  T}}\sim \mathcal{O}(m_{3/2})$ (cf.\ e.g.\ \cite{hep-ph/9710314}). The computation of their masses requires the knowledge of the soft-breaking sector and the mediation to the visible sector. For the sake of the argument, we assume SUSY breaking via gravity mediation by an unspecified hidden sector. Then, the superpotential consists of the observable and the hidden sector parts,
\begin{equation}
    W=W_{\mathcal O}+W_{\mathcal H},
\end{equation}
and we may consider the following  K\"{a}hler potential:
\begin{align}
     K&=K_{\mathcal H}+S^\dagger S +\phi^\dagger\phi+\frac{\kappa}{m_{Pl}^2}S^\dagger S\:\phi^\dagger\phi+\frac\alpha{m_{Pl}^2}(S^\dagger S)^2
     +\frac{\beta}{m_{Pl}^2}(\phi^\dagger\phi)^2
     \nonumber\\&
     +\frac{\Omega^\dagger \Omega}{m_{Pl}^2}\Big(\eta_SS^\dagger S+\eta_\phi \phi^\dagger\phi\Big)+...\:.
\end{align}
Here $\Omega$  stands for the SUSY breaking hidden sector field, and $\phi$ is the SM-singlet of the adjoint as defined in Eq.~\eqref{eq:branching24}. In this part of the calculation, we can ignore the contributions from the matter fields since we assume that $f_3=0$ for $N_i=0$.  
We then compute the soft breaking terms in the usual limit $m_{Pl}\rightarrow\infty$ while keeping $m_{3/2}$ fixed (cf.~\cite{cHohl} eq. (2.209)-(2.212)):
\begin{align}
    V&=\Big|f_1(\phi)- \varkappa M^2\Big|^2+S^\dagger S\sum_{(a)}\Big[\frac{\partial f_1}{\partial\phi_{(a)}}\frac{\partial(f_1^*)}{\partial\phi_{(a)}^*}\Big]_\phi
    +\lambda_\phi m_{3/2}^2 \phi^\dagger\phi+\lambda_Sm_{3/2}^2S^\dagger S \nonumber
    \\
    &+m_{3/2}\Big[\lambda_WW+\lambda_{\partial f_1}S\phi\frac{\partial f_1}{\partial\phi}+\lambda_3S(S^\dagger S)+c.c.\Big] . \label{eq:soft}
\end{align}
The coefficients $\lambda_i$'s are of order one and depend on $\vev{\Omega}/m_{Pl}$ and $\vev{\phi}/m_{Pl}$ .$(a)$ is an adjoint $SU(5)$-index. Once the complete hidden sector is specified, these coefficients become explicitly calculable. 
One condition for the minimum of $V$ is that its derivative $\partial_{S^\dagger}V$ with respect to $S^\dagger$ vanishes, which yields
\begin{align}
    \partial_{S^\dagger}V&=S\bigg(\sum_{(a)}\Big[\frac{\partial f_1}{\partial\phi_{(a)}}\frac{\partial(f_1^*)}{\partial\phi_{(a)}^*}\Big]_\phi+\lambda_Sm_{3/2}^2\Bigg)
    +m_{3/2}\Big[-\lambda_W^*F_S+\lambda_{\partial f_1}^*\phi\frac{\partial f_1^*}{\partial\phi^*}\Big]   
    \nonumber\\
    &+m_{3/2}\Big[\lambda_3S^2+2\lambda_3^*SS^\dagger\Big] \stackrel{!}{=} 0.\label{eq:vCo}
\end{align}
Since Eq.~\eqref{eq:vCo} is violated for $F_S=0$ and $\vev S=0$ (corresponding to $F_{\phi_{(a)}}=0$), we conclude that the soft breaking terms induce slight shifts to $\vev \phi$ and $\vev S$.
To estimate the masses of the octet and the triplet, we need to determine the order of $\vev S$. To this end, we first expand $f_1$ 
\begin{equation}
f_1(\phi)=\phi^2\bigg(\sum_{i\geq0}c_i\Big(\frac\phi\Lambda\Big)^i\bigg),   \label{eq:fexpansion} 
\end{equation}
which leads to
\begin{equation}
    \frac{\partial f_1}{\partial \phi}=\phi\bigg(\sum_{i\geq0}c_i(2+i)\Big(\frac\phi\Lambda\Big)^i\bigg).   
\end{equation}
For simplicity, we will replace $\phi$ with $M_\mathrm{GUT}$ in the following and shall ignore the shift induced on $\vev\phi$. Defining $\varepsilon_\Lambda\equiv M_\mathrm{GUT}/\Lambda$ and using the fact that $M_\mathrm{GUT}\gg m_{3/2}$, the vev of $S$ turns out to be 
\begin{equation}
\vev{S}\sim m_{3/2}\varepsilon_\Lambda^{-n}, \hspace{30pt} 
n\equiv\mathrm{min}\{\:i\:|c_i\neq0\}. \label{eq:Svev}
\end{equation}
Applying the expansion Eq.~\eqref{eq:fexpansion} and substituting the vev of $S$ from Eq.~\eqref{eq:Svev} to Eq.~\eqref{eq:fermionmass}, we finally obtain the masses of the triplets and the octets of order the gravitino mass,
\begin{align}
    \mathcal L&\supset S\Big[\tilde f_{1,\mathscr O}(\vev{\Phi})\tilde {\mathscr 
 O}\tilde{\mathscr  O}+\tilde f_{1,\mathscr T} (\vev{\Phi}) \tilde{\mathscr  T}\tilde{\mathscr  T}\Big]+H.c. \;\sim \;  m_{3/2}(\tilde {\mathscr O} \tilde{\mathscr O}+ \tilde{\mathscr T}\tilde{\mathscr T})+H.c.\:.
\end{align}
On the one hand these light relics represent interesting signatures of the underlying GUT theory, potentially testable at colliders. On the other hand, they tend to disturb the otherwise ``automatic'' gauge coupling unification in the MSSM.


\begin{figure}
    \centering
    \includegraphics[width=0.8\textwidth]{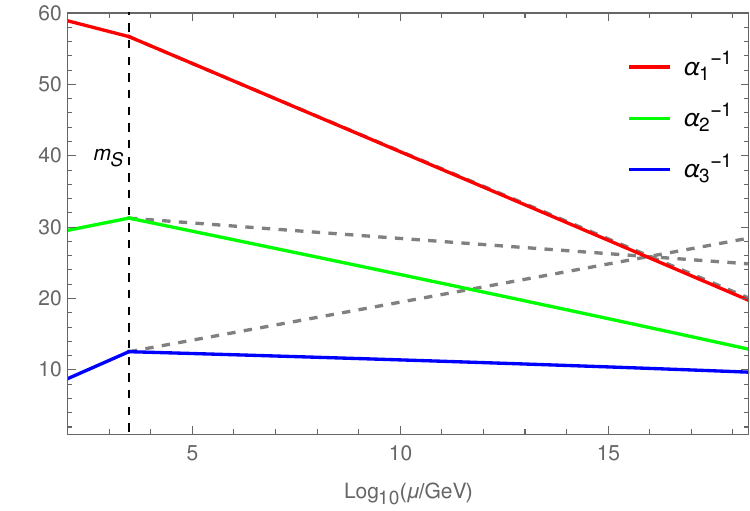}
    \caption{Running of the gauge couplings in the presence of SUSY scale triplets and octets. The dashed lines represent the successful gauge coupling unification in the MSSM scenario at around $2\times 10^{16}$ GeV. The SUSY scale is taken to be 3 TeV.}
    \label{fig:broken}
\end{figure}

\section{Gauge Coupling Unification in the Presence of Light Triplets and Octets}\label{sec:GCU}
With only extra light triplets and octets in the spectrum, in addition to the MSSM, the successful gauge coupling unification (GCU) is spoiled, as depicted in Fig.~\ref{fig:broken}. 
In order to restore GCU, we introduce additional conjugate pairs of chiral supermultiplets.
It is suggested in Refs.~\cite{1010.3657,1804.07619,1910.07554} that multiple copies of the following vectorlike pairs of chiral supersymmetric states may be used for restoring gauge coupling unification: 
\begin{align}
&\overline 5+5=
\left\{ \overline D(1,2)_{-1/2}\oplus\overline T(\overline 3,1)_{1/3}  \right\}
+
\left\{  D(1,2)_{1/2} \oplus T(3,1)_{-1/3} \right\},
\\
&\overline{10}+10=
\left\{ 
\overline E^c(1,1)_{-1}\oplus \overline  Q(\overline 3,2)_{-1/6}\oplus \overline U^c(3,1)_{2/3}
\right\}
+
\nonumber \\&\hspace{55pt}
\left\{ 
E^c(1,1)_1\oplus Q(3,2)_{1/6}\oplus  U^c(\overline 3,1)_{-2/3}
\right\}.
\end{align}
Since a complete GUT multiplet cannot affect the running of the gauge couplings, it is necessary to split the submultiples such that some components reside much below the GUT scale. However, this splitting, achieved in an \textit{ad hoc} way by adjusting the coefficients of some superpotential operators, 
introduces additional severe fine-tuning. For example, let us focus on a pair of vectorlike chiral supermultiplets $5+\overline 5$, and consider, e.g., the following terms in the superpotential: 
\begin{equation}
    \delta W^{d=3}\supset m_5 \overline 55 + \lambda_5 (5)^A(\Phi)_A^B(\overline 5)_B,
\end{equation}
with $A,B=1,...,5$ being fundamental $SU(5)$ indices. The above terms are sufficient to split the lepton-like components from the down-quark type components due to the relative Clebshes, when the parameters $m_5$ and $\lambda_5$ are finely adjusted. However, we would like to avoid such tuning. Therefore, we will propose a mechanism to realise the needed splitting.

\begin{figure}
    \centering
    \includegraphics[width=.9\textwidth]{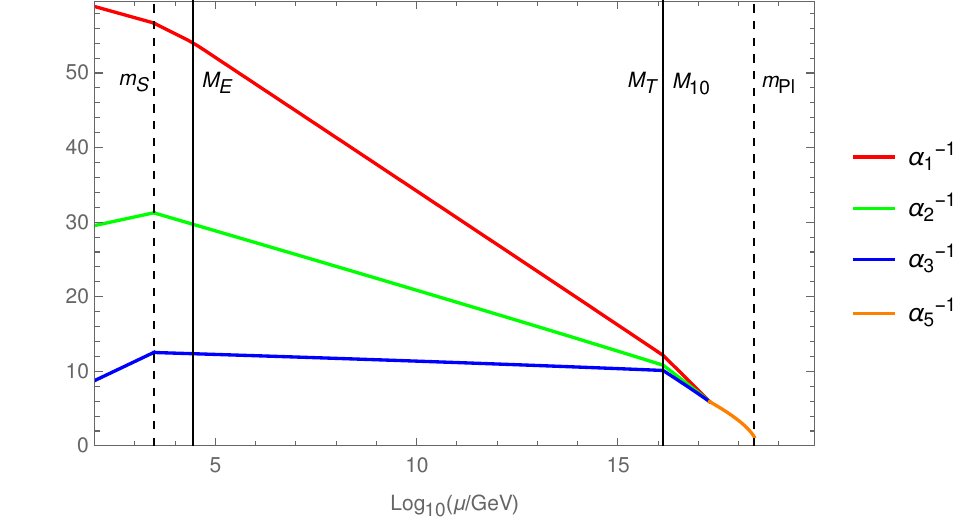}
    \caption{{ Running of gauge couplings at two-loop. As discussed in the text, the effects of the Yukawa couplings and threshold corrections are neglected. The SUSY scale is taken to be $3\,\mathrm{TeV}$, $m_{Pl}$ is the reduced Planck mass, and the rest of the mass scales are set with respect to the benchmark point of Table \ref{tab:bench}.}}
    \label{fig:gcu1}
\end{figure}

We  first discuss the restoration of gauge coupling unification within our scenario. For this purpose, we consider the following set of states (with $\alpha=1,2$) that live below the GUT scale:
\begin{multline}
    \overline D'(1,2)_{-1/2}+ D'(1,2)_{1/2}+ E^c_\alpha(1,1)_1+\overline E^c_\alpha(1,1)_{-1}\oplus \overline T_\alpha(\overline3,1)_{1/3}+T_\alpha( 3,1)_{-1/3}\\+Q_\alpha(3,2)_{1/6} +\overline Q_\alpha(\overline 3,2)_{-1/6}+ \overline U^c_\alpha(3,1)_{2/3}+ U^c_\alpha(\overline 3,1)_{-2/3}.
\end{multline}
The origin of these submulitplets will be discussed in the next subsections. 
We assume that $Q_\alpha+\overline Q_\alpha+ U_\alpha+\overline U_\alpha$ and $T_\alpha+\overline T_\alpha$ have masses of similar order, which we  denote by $M_{10}$ and $M_T$, respectively. Furthermore, we take $D'+ \overline D'$ to have SUSY scale masses, which will be justified in Sec.~\ref{sec:dmpm}. The mass scale $M_E$ of the singly charged states $E^c_\alpha+\overline E^c_\alpha$ is taken as a free parameter. Therefore, there are five parameters $\left(M_{10}, M_T, M_E, g_{5}, M_\mathrm{GUT}\right)$ to obtain gauge coupling unification and fit the low-scale ($M_Z$) measured values of  $\left(g_1, g_2, g_3\right)$ for a given SUSY-scale $m_{S}$. In Table~\ref{tab:bench}, we present an example (benchmark point) where successful gauge coupling unification. The corresponding $\beta$-functions are given in Appendix~\ref{a:gcu}, and the restoration of the gauge coupling unification is depicted  in Fig. \ref{fig:gcu1}.

Note that, for the benchmark point, $M_E$ is less than one order of magnitude above the SUSY scale. This is partially due to the fact that the one-loop $\beta$-functions of the octet, the triplets, the electron-like states, and the lepton-like doublets (see Appendix \ref{a:gcu}) add up to
\begin{equation}
    b_a^{\mathscr T}+b_a^{\mathscr O}+2b_a^{D'}+4b_a^{E^c_\alpha}=\begin{pmatrix}3\\3\\3\end{pmatrix},
\end{equation}
thus approximately restoring automatic GCU of the MSSM. This was already pointed out in Ref.~\cite{1010.3657} in a similar context and in Refs.~\cite{0807.3055,0910.2732,0910.3020} in the context of the little hierarchy problem.

The presence of a ``desert'' between $M_E$ and $M_{T,10}$ suggest that in a such a setup only two scales are of importance, namely the SUSY scale $m_S$ and the GUT scale $M_\mathrm{GUT}$.

\begin{table}
    \centering
    \begin{tabular}{llllll}\hline\hline
     \multicolumn{6}{l}{Benchmark Point}\\\hline
       $m_{S}$&$g_{5}$&$M_\mathrm{GUT}$&$M_E$&$M_T$&$M_{10}$\\
         $3.00\times 10^3$&$1.45$&$1.90\ee{17}$&$2.81\ee{4}$&$1.30\ee{16}$&$1.33\ee{16}$\\
         \hline\hline
\end{tabular}
    \caption{ A benchmark point for achieving gauge coupling unification at two-loop; see text for details. All mass scales are given in GeV.  }
    \label{tab:bench}
\end{table}

\subsection{Double Missing Partner Mechanism}\label{sec:dmpm}

We first elude the origin of $T_\alpha+\overline T_\alpha$ and $D'+\overline D'$, and how the splitting is achieved via the MPM~\cite{1982PhLB..115..380M,HUTP-82/A014}
or rather its extended version, namely, the DMPM~\cite{hep-ph/9406417,hep-ph/9611277,1405.6962}.  To that end, we introduce, on top of the usual content of an SUSY-$SU(5)$ GUT, two pairs of $50+\overline{50}$ Higgs multiplets, namely, $50_X+ \overline{50}_X+ 50_Y+ \overline{50}_Y$, as well as one pair of $5'+\overline 5'$ Higgs multiplet, and consider the following superpotential: 
\begin{align}
    W_{DMPM}&=\frac{\alpha_2}{\Lambda}5'(\Phi^2)_{75}\overline{50}_X+\frac{\alpha_4}{\Lambda}\overline 5'(\Phi^2)_{75}50_Y\nonumber\\
    &+\frac{\alpha_1}{\Lambda}5_H(\Phi^2)_{75}\overline{50}_Y+\frac{\alpha_3}{\Lambda}\overline 5_H(\Phi^2)_{75}50_X\label{eq:dmpm}\\
    &+M_X 50_X\overline{50}_X+M_Y 50_Y\overline{50}_Y,\nonumber\\
    &+\delta W_\mathrm{eff}\nonumber,
\end{align}
where 
\begin{equation}
     \delta W_\mathrm{eff}=\mu_H 5_H\overline 5_H+\mu'5'\overline 5'\;, \label{eq:mudmpm}
\end{equation}
and
\begin{equation}
   m_{S}\sim\mu_H\lesssim\mu'\ll\alpha_iv_{24}^2/\Lambda\ll M_{X,Y}\sim m_{Pl},\label{eq:dmpmscales}
\end{equation}
where $v_{24}$ is defined in the Appendix~\ref{a:40b} (see Eq.~\eqref{v24}). 
The smallness of $\mu_H$ and $\mu'$ could be explained similarly as the smallness of the MSSM $\mu$-term, e.g.\ by introducing higher dimensional operators \textit{\`a la} Froggatt-Nielsen \cite{CERN-TH-2519} or by realising a Giudice-Masiero-like mechanism \cite{1988PhLB..206..480G} and thus tying the doublet masses to the SUSY scale. Regarding the latter option, one may e.g.\ consider the following additional terms in the K\"ahler potential,
\begin{equation}
    \delta K=\frac{Z_H^\dagger}{\Lambda_5}\overline 5_H5_H+\frac{Z^{\prime\dagger}}{\Lambda_5}\overline 5^\prime5^\prime+c.c.,\label{eq:dK1}
\end{equation}
where $Z_H$ and $Z^\prime$ obtain $F$-terms in the process of SUSY breaking. Plugging in these F-terms  and integrating over $d^2\bar \theta$ superspace leads to the above listed effective superpotential terms, with $\mu_H$ and $\mu'$ of the order of the SUSY scale. However one should keep in mind that SUSY is broken and there are additional terms in the Lagrangian.   

Note that from a model-building perspective it can be favourable to generate $M_{X,Y}$ via $SU(5)$ singlets getting a Planck-scale vev. Defining
\begin{align}
    &T_H+ \overline T_H\subset5_H+\overline5_H, \;\;
    T'+ \overline T'\subset5'+\overline5', 
    \nonumber \\ 
    &T_X+\overline T_X\subset50_X+\overline{50}_X, \;\; 
    T_Y+\overline T_Y\subset50_Y+\overline{50}_Y,
\end{align}
once $\Phi$ acquires a vevs, one is left with the following superpotential for the doublets and triplets:
\begin{align}
    W^{eff}_{DMPM}&=\begin{pmatrix}T_H&T'&T_X&T_Y\end{pmatrix}
   \underbrace{ \begin{pmatrix}
    \mu_H&0&0&\alpha_1v_{24}^2/\Lambda\\
    0&\mu'&\alpha_2v_{24}^2/\Lambda&0\\
    \alpha_3 v_{24}^2/\Lambda&0&M_X&0\\
    0&\alpha_4v_{24}^2/\Lambda&0&M_Y
    \end{pmatrix}}_{\equiv\mathcal{M}_T}
    \begin{pmatrix}\overline T_H\\\overline T'\\\overline T_X\\\overline T_Y\end{pmatrix}\nonumber\\
    &+\begin{pmatrix}H_u&D' \end{pmatrix}
    \underbrace{ \begin{pmatrix}
    \mu_{H}&0\\
    0&\mu'
    \end{pmatrix}}_{\equiv\mathcal{M}_D}
    \begin{pmatrix}H_d\\\overline D' \end{pmatrix}. \label{eq:DMPMeff}
\end{align}

The states from the $50$s that are not present in Eq.~\eqref{eq:DMPMeff}, all get masses of order $M_{X,Y}$. Labelling the mass eigenstates of $\mathcal M_T$ by $T_m+\overline T_m,\,m=1,...,4$, we expect $T_{3,4}+\overline T_{3,4}$ to be heavy, i.e., of order $M_{X,Y}$, while $T_{1,2}+\overline T_{1,2}$, mainly originating from $5'+\overline5'$ and $5_H+\overline 5_H$, intermediate-scale mass $M_T<M_\mathrm{GUT}$. By neglecting $\mu_H$ and $\mu'$, one obtains 
\begin{equation}
    (\mathcal M_T)_1(\mathcal M_T)_2(\mathcal M_T)_3(\mathcal M_T)_4=\mathrm{det}(\mathcal M_T)=\overline{\alpha}^4\Big(\frac{v_{24}^2}{\Lambda}\Big)^4 ,
\end{equation}
where, we have defined $\overline\alpha$ as the geometric mean of $\alpha_1,\dots,\alpha_4$ and $(\mathcal M_T)_m$ are the mass eigenvalues of $\mathcal M_T$. Therefore, we expect $M_T$ of order
\begin{equation}
    M_T\sim \overline\alpha^2\frac{M_\mathrm{GUT}^4}{\Lambda^2M_{X,Y}}. \label{eq:mass50}
\end{equation}
Eq.~\eqref{eq:mass50} relates $M_\mathrm{GUT}$ and $M_T$ to $M_{X,Y}$ and the messenger scale $\Lambda$. One thus has to check whether it is possible to accommodate $M_{X,Y}$ and $\Lambda$ below $m_{Pl}$, and inspect their implications for perturbativity. However in the case of the values of Table \ref{tab:bench} this possible.

Now, comparing Eq.~\eqref{eq:DMPMeff}, Eq.~\eqref{eq:dmpmscales}, and Eq.~\eqref{eq:mass50}, we thus conclude that the doublets remain light, whereas the triplets acquire superheavy masses somewhat below the GUT scale. Additionally,  this splitting suppresses dimension five nucleon decay.  Since the MSSM species only couple to $T_H+ \overline T_H$,  the  effective triplet mass, relevant for nucleon decay, is $M_T^{eff}\equiv\big((\mathcal M_T^{-1})_{11}\big)^{-1}$, which is given by~\cite{hep-ph/9611277}:
\begin{equation}
    M_T^{eff}\sim \frac{v_{24}^8\overline\alpha^4}{\mu'M_XM_Y\Lambda^4}\approx\frac{M_T^2}{M_{D'}}.
\end{equation}
Then the decay width of the proton can be approximately written as (cf.~\cite{0912.5375} and references therein), 
\begin{equation}
    \Gamma_p\approx|\mathcal A|^2m_p^5, \quad\quad\mathcal A_{D=5}\approx \frac{g_5^2}{(4\pi)^2}\frac{y_u y_d}{M_T^{eff}m_S}.
    \end{equation}
A crude estimate for the values of Tab.~\ref{tab:bench} then  yields a proton lifetime of 
\begin{equation}
    \tau_p\sim10^{56}\:\mathrm{ years},
\end{equation}
which is far out of reach of any current/planned future experiment thus rendering the dimension 6 contributions dominant.

\subsection{Generalised Missing Partner Mechanism}
In this subsection, by extending the concept of the Missing Partner Mechanism, we show how to obtain light $E_\alpha+\overline E_\alpha$ states naturally. To realise this, we introduce  two pairs of conjugate chiral 40-plets $40_I+\overline{ 40}_I$ (with $I=1,2$) and two pairs of conjugate $10$'s. Thus, including the SM fermions, we now have five families of $10_a$, where $a=1,...5$ and two generations of $\overline{10}_I$ with $I=1,2$. The superpotential we consider takes the following form:  
\begin{align} 
W_{EMPM}&=(Y_{40})_{IJ}40_I\Phi \overline{10}_{J}+(Y_{\overline{40}})_{Ia}\overline{40}_I\Phi{10}_{a} \nonumber\\
&+(M_{40})_{IJ}\overline{40}_I40_J+\delta W_\mathrm{eff},\label{eq:Wempm}
\end{align} 
where 
\begin{equation}
    \delta W_\mathrm{eff}=(m_E)_{Ia}\overline{10}_I10_a.
\end{equation}
As in equations \eqref{eq:dmpm} and \eqref{eq:mudmpm}, $\delta W_\mathrm{eff}$ should be understood as in principle forbidden, only to be reintroduced via some mechanism as to justify 
\begin{equation}
    m_{3/2}\lesssim(m_E)_{IJ}\ll v_{24}(Y_{40,\overline{40}})_{IJ}\ll (M_{40})_{IJ},
\end{equation}
where it may, from a model building perspective, be favourable to generate $M_{40}$ via the vev of a singlet superfield. Note that in the case of a Giudice-Masiero-like mechanism,  the slight difference between $m_E$ and $\mu^\prime/\mu_H$ can originate form various sources. Writing
\begin{equation}
    \delta K\subset\frac{Z^\dagger_{10}}{\Lambda_{10}}10_a(y_{Z,E})_{aI}\overline{10}_I,
\end{equation}
and comparing it to \eqref{eq:dK1}, the order of magnitude difference between $m_S$ and $M_E$ in Fig.~\ref{fig:gcu1} can be explained e.g.\ from $\vev{Z_{10}}>\vev{Z_5}$, $\vev{F_{Z_5}}>\vev{F_{Z_{10}}}$, or due to $\Lambda_5>\Lambda_{10}$ arising due to a different origin of the different operators. The representation
$\overline{{40}}$ of $SU(5)$ decomposes under $G_{SM}$ as 
\begin{align}
\overline{{40}}&=\overline Q^{40}( \overline{ 3}, 2)_{-1/6}\oplus \overline U^{c,40}( 3, 1)_{2/3}\oplus\Pi({ 3}, 3)_{2/3}
\nonumber\\&\oplus\Sigma( 8, 1)_{-1}\oplus\Psi( 6, 2)_{-1/6}\oplus\Delta( 1, 2)_{3/2}\label{eq:40}.
\end{align}
From the above decomposition, it is interesting to note that the coloured submultiplets residing in  ${10}$ have conjugated partners in $\overline{40}$, which however, is not true for the state $\overline E^c$. This offers the possibility of getting  $E^c+\overline E^c$ states naturally light by generalising the MPM scheme. Therefore, once $\Phi$ acquires a vev, the operator ${10}_{a}\Phi\overline{{40}}_I$ only generates masses for the quark-like states of $\overline{40}$ and $10$, i.e.\ 
\begin{equation}
\varepsilon_{ABCDE}({10}_{a})^{[AB]}(\Phi)^C_F(\overline{{40}}_I)^{[DE]F} \supset \;
\underbrace{ \frac{2\sqrt{10}v_{24}}3 }_{\equiv v^\prime_{24}}(\overline Q_I^{40}Q^{10}_{a}+\overline U_I^{c,40}U^{c,10}_{a}).
\end{equation}
The explicit decomposition of $\overline{40}$ and our conventions are presented in Appendix \ref{a:40b}. Once $\Phi$ acquires a vev, we are left with the following mass terms in the superpotential for the submultiplets of  $10_{a}\oplus\overline{10}_{I}$ and $40_{I}\oplus\overline{40}_{I}$:
\begin{align}
W_{mass}&=
\begin{pmatrix}(\overline Q^{10})^T&(\overline{Q}^{40})^T\end{pmatrix}
\underbrace{
\begin{pmatrix}m_E&v_{24}^\prime Y_{40}^T\\v^\prime_{24}Y_{\overline{40}}&M_{40}\end{pmatrix}
}_{\equiv\mathcal M_Q}
\begin{pmatrix}Q^{10}\\Q^{40} \end{pmatrix}\nonumber\\
&+
\begin{pmatrix}(\overline U^{c,10})^T&(\overline{U}^{c,40})^T\end{pmatrix}
\begin{pmatrix}m_E&v^\prime_{24}Y_{40}^T\\v^\prime_{24}Y_{\overline{40}}&M_{40}\end{pmatrix}
\begin{pmatrix}U^{c,10}\\U^{c,40} \end{pmatrix}
\label{eq:Qmasseff}\\
&+(\overline E^c)^{T} m_E E^c+\Pi ^TM_{40}\overline\Pi+\Sigma ^TM_{40}\overline\Sigma+\Psi ^TM_{40}\overline\Psi+\Delta ^TM_{40}\overline\Delta.\nonumber
\end{align}
Note that $\mathcal M_Q$ is in fact a $4\times7$ matrix while $m_E$ is a $2\times5$ one.  Without loss of generality we can take the following form for $m_E$ and $ Y_{\overline{40}}$:
\begin{align}
    m_E&=
    \begin{pmatrix}
        \eta_{11}&\eta_{21}&\eta_{31}&\eta_4&0\\
        \eta_{12}&\eta_{22}&\eta_{32}&0&\eta_5
    \end{pmatrix},\nonumber\\
    Y_{\overline{40}}&=
    \begin{pmatrix}
        y^{\overline{40}}_{11}& y^{\overline{40}}_{12}& y^{\overline{40}}_{13}& y^{\overline{40}}_{4}& 0\\
         y^{\overline{40}}_{21}& y^{\overline{40}}_{22}& y^{\overline{40}}_{23}&y^{\overline{40}}_{5}& y^{\overline{40}}_{6} 
    \end{pmatrix},\label{eq:texture}
\end{align}
with $\eta_4,\,\eta_5,\, y^{\overline{40}}_{4},\,y^{\overline{40}}_{5}$ and $y^{\overline{40}}_{6}$ being real and positive. Performing re-phasings and rotations, one can bring $Y_{\overline{40}}$ and $m_E$ into the following forms:
\begin{align}
   m_EP(\varphi^E_{11},\varphi^E_{12},\varphi^E_{13})R_{14}(\theta^E_{14})R_{24}(\theta^E_{24})R_{34}(\theta^E_{34})\times&\nonumber\\
    \times P(\varphi^E_{21},\varphi^E_{22},\varphi^E_{23})R_{15}(\theta^E_{15})R_{25}(\theta^E_{25})R_{35}(\theta^E_{35})&=
    \begin{pmatrix}
        0&0&0&M_{E,1}&0\\
        0&0&0&M_{E,2}&M_{E,3}
    \end{pmatrix}\\
    &\equiv m_EU_E , \nonumber\\
    Y_{\overline{40}}P(\varphi^{\overline{40}}_{11},\varphi^{\overline{40}}_{12},\varphi^{\overline{40}}_{13})R_{14}(\theta^{\overline{40}}_{14})R_{24}(\theta^{\overline{40}}_{24})R_{34}(\theta^{\overline{40}}_{34})\times&\nonumber\\
    \times P(\varphi^{\overline{40}}_{21},\varphi^{\overline{40}}_{22},\varphi^{\overline{40}}_{23})R_{15}(\theta^{\overline{40}}_{15})R_{25}(\theta^{\overline{40}}_{25})R_{35}(\theta^{\overline{40}}_{35})&=
    \begin{pmatrix}
       0&0&0&y_1^{\overline{40}}&0\\
        0&0&0&y_2^{\overline{40}}&y_3^{\overline{40}}
    \end{pmatrix}\\
    &\equiv Y_{\overline{40}}U_{\overline{40}} ,\nonumber
\end{align}
where,
\begin{equation}
    p(\alpha,\beta,\delta)=\mathrm{diag}(e^{i\alpha},e^{i\beta},e^{i\delta},1,1,1,1),
\end{equation}
and
\begin{equation}
    R_{ij}(\theta)=
    \begin{pmatrix}
        \cos \theta&\sin\theta\\
        -\sin\theta&\cos\theta
    \end{pmatrix},
\end{equation}
stands for rotation in the $i-j$ plane. Defining 
\begin{equation}
    \begin{pmatrix}e^c_1\\e^c_2\\e^c_3\\{E^{c}_1}'\\{E^c_2}'\end{pmatrix}=U_E^\dagger\begin{pmatrix}E^c_1\\E^c_2\\E^c_3\\E^c_4\\E^c_5\end{pmatrix},
    \;
    \begin{pmatrix}u^c_1\\u^c_2\\u^c_3\\{U^{c}_1}'\\{U^c_2}'\end{pmatrix}=U_{\overline{40}}^\dagger\begin{pmatrix}U^{c,10}_1\\U^{c,10}_2\\U^{c,10}_3\\U^{c,10}_4\\U^{c,10}_5\end{pmatrix},
    \;\text{and}\;
    \begin{pmatrix}q_1\\q_2\\q_3\\{Q^{c}_1}'\\{Q^c_2}'\end{pmatrix}=U_{\overline{40}}^\dagger\begin{pmatrix}Q^{10}_1\\Q^{10}_2\\Q^{10}_3\\Q^{10}_4\\Q^{10}_5\end{pmatrix},
\end{equation}
we may rewrite \eqref{eq:Qmasseff} in terms of the new fields:
\begin{align}
W_{mass}&\supset
\begin{pmatrix}(\overline Q^{10})^T&(\overline{Q}^{40})^T\end{pmatrix}
\mathcal M_Q
\begin{pmatrix}U_{\overline{40}}U_{\overline{40}}^\dagger Q^{10}\\Q^{40} \end{pmatrix}\nonumber\\
&+
\begin{pmatrix}(\overline U^{c,10})^T&(\overline{U}^{c,40})^T\end{pmatrix}
\mathcal M_Q
\begin{pmatrix}U_{\overline{40}}U_{\overline{40}}^\dagger U^{c,10}\\U^{c,40} \end{pmatrix}
\\
&+(\overline E^c)^{T} m_EU_EU_E^\dagger E^c\nonumber\\
&=\begin{pmatrix}(\overline Q^{10})^T&(\overline{Q}^{40})^T\end{pmatrix}
\underbrace{
\begin{pmatrix}
    \eta&Y^T_{40}v^\prime_{24}\\
    \tilde Y_{\overline{{40}}}v^\prime_{24}&M_{40}
\end{pmatrix}
}_{\mathcal M_Q'}
\begin{pmatrix}Q'\\Q^{40} \end{pmatrix}\nonumber\\
&+\begin{pmatrix}(\overline U^{c,10})^T&(\overline{U}^{c,40})^T\end{pmatrix}
\begin{pmatrix}
    \eta&Y^T_{40}v^\prime_{24}\\
    \tilde Y_{\overline{{40}}}v^\prime_{24}&M_{40}
\end{pmatrix}
\begin{pmatrix} {U^{c}}'\\U^{c,40} \end{pmatrix}\nonumber\\
&+(\overline E^c)^{T} \mathcal M_E{E^c}'+(\overline Q^{10})^T\eta'q+(\overline U^{c,10})^T\eta'u^c\nonumber,
\end{align}
where
\begin{equation}
\tilde Y_{\overline{40}}\equiv
\begin{pmatrix}
   y_1^{\overline{40}}&0\\
    y_2^{\overline{40}}&y_3^{\overline{40}} 
\end{pmatrix},\quad
\mathcal M_E\equiv\begin{pmatrix}
M_{E,1}&0\\
M_{E,2}&M_{E,3}\end{pmatrix},\quad\text{and}\quad
(\eta'|\eta)\equiv m_EU_{\overline{40}} . 
\end{equation}
Note that $e^c_i$ are completely massless while even though $u^c_i$ and $q_i$ are not exactly the massless eigenstates the error is of order $\mathcal O(M_{E,i}/M_\mathrm{GUT})$ and thus negligible. Those states are to be identified with the ones of the MSSM. We label the mass eigenstates of $\mathcal M'_Q$  by $Q_m\oplus \overline Q_m$ and $U_m\oplus \overline U_m$ ($m=1,...,4$), respectively. Due to the strong diagonal hierarchy, we expect $Q_{3,4}\oplus \overline Q_{3,4}$ and $U_{3,4}\oplus \overline U_{3,4}$ to originate mainly from $40_I\oplus\overline{40}_I$ with similar masses to the other states coming from the same multiplets. On the other hand, $Q_{1,2}\oplus \overline Q_{1,2}$ and $U_{1,2}\oplus \overline U_{1,2}$ mainly originate from $10_{a}\oplus\overline{10}_{I}$ and have somewhat smaller masses. Labelling the masses of theses eigenstates by $(\mathcal M_Q')_m$ they can be estimated by considering the determinant of $\mathcal M_Q'$:
\begin{align}
    \mathrm{det}(\mathcal M_Q')&=(\mathcal M_Q')_1(\mathcal M_Q')_2(\mathcal M_Q')_3(\mathcal M_Q')_4\nonumber\\
    &=\mathrm{det} (M_{40}) \det(\eta-v_{24}^{\prime2}Y_{40}^TM_{40}^{-1}\tilde Y_{\overline{40}}) \nonumber\\
    &\sim v_{24}^{\prime4}, \label{eq:det}          
\end{align}
where in the last step, we have neglected $\eta$, and assumed that $\det(Y_{40,\overline{40}})\sim \mathcal O(1)$ \footnote{Additionally one also has to assume that $M_{40}$ is non-singular.}. Utilising $(\mathcal M)_{3,4}$ are of the same order as the entries of $M_{40}$, say $(\mathcal M)_{3,4}\sim m_{40}$,  we deduce that
\begin{equation}
   M_{10}\sim (\mathcal M)_{1,2}\sim  M_\mathrm{GUT}\bigg(\frac{M_\mathrm{GUT}}{m_{40}}\bigg). \label{eq:mass40}
\end{equation}
Thus \eqref{eq:mass40}, together with \eqref{eq:mass50} and the suppressed dimension-full parameters of $\delta W_\mathrm{eff}$ give rise to a ``desert'' as seen in Fig.~\ref{fig:gcu1} and Tab.~\ref{tab:bench}.

Note that Eq.~\eqref{eq:mass40} relates $M_{10}$ and $M_\mathrm{GUT}$ to $M_{40}$. Therefore, one has to check whether values of $M_{10}$ and $M_\mathrm{GUT}$ are consistent with the eigenvalues of $M_{40}$ being below $m_{Pl}$, and determine the scale up to where the unified gauge coupling remains perturbative (ideally up to $m_{Pl}$). 
For the benchmark point in Tab.~\ref{tab:bench}, this is satisfied.

\section{Fermion Masses}\label{sec:FM}

In the previous section, to restore the gauge coupling unification, we have introduced vectorlike fermionic states (and of course their scalar partners) with quantum numbers identical to the SM fermions, i.e., $ e^c, q, u^c$. It is interesting to note that these same states can be utilised to correct the wrong mass relations between the charged leptons and the down-type quarks due to mixing effects. For previous works on correcting fermion mass relation by employing vectorlike states, see, e.g., Refs.~\cite{1985PhLB..150..177B,hep-ph/9503215,hep-ph/9511446,hep-ph/9512389,hep-ph/9904249,hep-ph/0305090,0710.0581,0907.3400,1207.6388,1401.6870,1910.09008,2301.00809,Antusch:2023mqe}.

In addition to the superpotential terms given in Eq.~\eqref{eq:Wempm}, the following Yukawa interactions contribute to the fermion masses, 
\begin{equation}
W_{Y}=\frac18(Y_u)_{ab}10_a10_b5_H+(Y_d)_{ia}\overline 5_i10_a\overline 5_H+W_\nu,
\end{equation}
where $Y_u$ is a $5\times 5$ matrix, and $Y_d$ is a $3\times 5$ matrix. Moreover, the part $W_\nu$ provides neutrino masses, which we do not specify in this work. On top of the textures of \eqref{eq:texture} we may assume without loss of generality
\begin{equation}
    Y_d=
    \begin{pmatrix}
        y_1&0&0&0&0\\
        0&y_2&0&0&0\\
        0&0&y_3&0&0
    \end{pmatrix},
\end{equation}
with real positive entries only. The MSSM Yukawas are now easily determined by deriving  how the light states couple to $H_d$ and $H_u$. We use R-L convention in the following, i.e.\ 
\begin{align}
    W^{MSSM}&=(Y_d^{MSSM})_{ij}d^c_i\ H_d\cdot q_j-(Y_u^{MSSM})_{ij}u_i^c\ H_u\cdot q_j\nonumber\\&+(Y_e^{MSSM})_{ij}e^c_i\ H_d\cdot \ell_j+\mu^{MSSM}H_u\cdot H_d \:.
\end{align}
The $SU(2)$ contractions are defined by $\psi\cdot\xi\equiv\varepsilon_{ab}\psi^a\xi^b$ with $\varepsilon$ the two dimensional Levi-Civita tensor\footnote{We take $\varepsilon_{12}=1$.}. The MSSM couplings are given by (see Appendix \ref{a:40b} for our $SU(5)$ conventions)
\begin{equation}
    (Y_d^{MSSM}|*)\equiv Y_dU_{\overline{40}},\quad(Y_e^{MSSM\;T}|*)\equiv Y_dU_E,\quad\mu^{MSSM}=-\mu_H\:,
\end{equation}
and
\begin{equation}
    \begin{pmatrix}
        Y_u^{MSSM}&*\\
        *&*
    \end{pmatrix}\equiv U_{\overline{40}}^T Y_u U_{\overline{40}}\;.
\end{equation}
Thus $Y_u^{MSSM}$ is a generic $3\times3$ matrix. Explicit computation for $Y_d^{MSSM}$ yields
\begin{equation}
    Y^{MSSM}_d=
    \begin{pmatrix}
       y_1e^{i(\varphi^{\overline{40}}_{11}+\varphi^{\overline{40}}_{21})}c^{\overline{40}}_{14}c^{\overline{40}}_{15}&\mathcal O(y_1)&\mathcal O(y_1)\\
       0&y_2e^{i(\varphi^{\overline{40}}_{12}+\varphi^{\overline{40}}_{22})}c^{\overline{40}}_{24}c^{\overline{40}}_{25}&\mathcal O(y_2)\\
       0&0&y_3e^{i(\varphi^{\overline{40}}_{13}+\varphi^{\overline{40}}_{23})}c^{\overline{40}}_{34}c^{\overline{40}}_{35}
    \end{pmatrix},
\end{equation}
and analogously for the charged leptons
\begin{equation}
    Y^{MSSM}_e=
    \begin{pmatrix}
       y_1e^{i(\varphi^{\eta}_{11}+\varphi^{\eta}_{21})}c^{\eta}_{14}c^{\eta}_{15}&0&0\\
       \mathcal O(y_1)&y_2e^{i(\varphi^{\eta}_{12}+\varphi^{\eta}_{22})}c^{\eta}_{24}c^{\eta}_{25}&0\\
       \mathcal O(y_1)&\mathcal O(y_2)&y_3e^{i(\varphi^{\eta}_{13}+\varphi^{\eta}_{23})}c^{\eta}_{34}c^{\eta}_{35}
    \end{pmatrix}.
\end{equation}
These above matrices contain enough free parameters to correct the mismatch between down-type quark and charged lepton masses as required by the experimental data.

\section{Phenomenology of Light Relics}\label{sec:pheno}

Light relic states, with masses around the SUSY scale, are a characteristic feature of the proposed scenario. In the following, we discuss present constraints on them and some of the observational signatures.

We note that within our scenario, the global $U(1)_R$ symmetry is broken by the SUSY breaking sector, which may leave a discrete $Z_2$ symmetry unbroken that can serve as R-parity. This leftover symmetry can stabilize the lightest supersymmetric particle (LSP), which is then a suitable DM candidate. 
If the LSP does not provide the whole DM abundance, within our considered scenario there could be additional stable neutral states that may contribute to the DM relic density, as will be discussed below.  

We now discuss the phenomenology of the light relics of our scenario:

\vspace{0.5cm}
\textbf{1.\;Colour Octet:}--
At the LHC, colour octets can be efficiently pair-produced via QCD interactions. They hadronize prior to decay, forming so-called R-hadrons~\cite{Farrar:1978xj} that are bound states composed of the octet and light quarks, antiquarks, and gluons. 
In our scenario, they are expected to be very long-lived (with details depending on the specific model).\footnote{An exception would be if there is a term which leads to a mixing between the octets and the gluinos. In this case they could decay rather fast into two gluinos (or into a gluon and a gluino in the case of the fermionic superpartner). Such terms could be achieved by introducing a gauged $U(1)$ in the hidden sector with non-vanishing D-term and an operator of the form 
$
\int d^2\theta \mathcal W^\alpha_{U(1)}\tr(\mathcal W_{SU(5),\alpha}\Phi).
$
See~\cite{1982PhRvD..26.3661P,1984PhLB..148..317J,1990PhRvL..65.2939H,hep-ph/9205227,hep-ph/9903365,hep-ph/0206096} for literature on such soft-terms.}
The collider signature of such coloured long-lived particles (CLLPs) are missing $E_T$ and a track in the calorimeter due to the small energy deposit. Non-observation at the LHC so far leads to a $95\%$CL bound of about $2$ TeV \cite{2205.06013} for stable gluino R-hadrons.  We can assume constraints on the (fermionic and scalar) colour octet relics from the adjoint of similar order.

Regarding cosmological constraints, the CLLPs confine into R-hadrons below the confinement temperature ($T_c\sim 200$ MeV). Subsequently, bound states of them are formed due to an unsuppressed rate of the capture process. These bound states with initially large angular momentum, $L$, lose energy and angular momentum by emitting pions and/or photons and finally decay. 
To avoid conflict with observations, this decay should happen before big bang nucleosynthesis (BBN). For the case of R-hadrons associated with gluinos, assumed to dominantly lose energy via loop-suppressed photon processes, the lifetime can be estimated as \cite{Kang:2006yd}
\begin{align}
\tau\sim \frac{4\pi m^6}{\alpha^2 \Lambda^7_\mathrm{had}}  \left( \frac{T_B}{\Lambda_\mathrm{had}} \right)^{7/3}  \:.
\end{align}
Using $\Lambda_\mathrm{had}\sim$ GeV, $T_B\sim T_c$,  and standard cosmology, the decay occurs before BBN for $m\lesssim 2.5$ TeV. Hence, colour octet masses $\gtrsim 2.5$ TeV are ruled out by (standard) cosmology. 
However, this bound might be relaxed when the loss of energy of the bound states via pions is included/allowed (which has not been considered in Ref.~\cite{Kang:2006yd}).

In summary, there is currently an interesting window of octet masses between about 2 TeV and 2.5 TeV which is untested. A discovery of a CLLP signal at future collider searches within this mass window could provide a smoking gun signal of the scenario.\footnote{Of course, the situation would change in case of a non-standard cosmological history, e.g.\ if there is late-time entropy production to sufficiently dilute the R-hadrons. Then the cosmological bounds could be much weaker, and also heavier colour octets are possible.}

\vspace{0.5cm}
\textbf{2.\;Weak Triplet and Doublets:}--
In the minimal version of our model, the weak doublets and triplets do not interact with SM fermions/scalars, they have only electroweak gauge interactions. In such a scenario, we refer to them as ``inert'' and they mimic the minimal dark matter candidates, as considered in Ref.~\cite{hep-ph/0512090,0706.4071,0903.3381}. However, additional interactions with the SM sector can be induced by allowing higher dimensional terms (e.g.\ by additional Giudice-Masiero K\"ahler potential terms generating a mixing with the other doublets). 

Due to only gauge interactions, as for the minimal DMs,  at the tree-level, the neutral as well as the charged components remain degenerate in mass. However, loop corrections involving gauge bosons typically make the charged components slightly heavier than the neutral one~\cite{hep-ph/0512090}. 
For the case of weak doublet (triplet), this mass splitting is of order $\Delta m\sim 350\; (166)$ MeV. 

To achieve the correct DM relic abundance, the mass of the neutral component of the doublet needs to be $m_\mathrm{DM}\sim 1100 \;(540)$ GeV for the case of fermionic (scalar) DM~\cite{hep-ph/0512090}. Since these doublets have non-zero hypercharge, they have vectorlike interactions with the $Z$-boson, leading to a tree-level spin-independent elastic cross-section, a few orders of magnitude above the current direct detection bounds. This stringent constraint from dark matter direct detection can be avoided if additional interactions are introduced. We assume such terms in the following.

On the other hand, a weak triplet with zero hypercharge is a viable candidate even with minimal interactions. Reproducing DM relic density demands the DM mass to be $\sim 2000 \;(2400)$ GeV for a scalar (fermion) triplet DM~\cite{hep-ph/0512090}. Since non-perturbative effects, such as Sommerfeld enhancement and bound-state effects, increase the annihilation cross-section, a slightly higher DM mass is required to satisfy the relic abundance, e.g., a fermionic DM mass of $\sim 2700$ GeV does the job~\cite{Mitridate:2017izz}.  Note, however, that since we may have multiple candidates for dark matter, each of these states must have masses below the numbers quoted above not to overproduce the total dark matter relic abundance.  

Of course, as commented above, these cosmological constraints could also be avoided in case of a non-standard cosmology scenario, e.g.\ if there is late-time entropy production that sufficiently dilutes the relic densities. Moreover, all cosmological bounds can be removed/relaxed by including higher dimensional terms to allow these states to mix with the visible sector. 

Note that both the charged and the neutral components of the weak triplet and the weak doublets can get efficiently produced at colliders via Drell-Yan processes due to their electroweak interactions. The small mass splitting between the DM components, as aforementioned, makes charged components long-lived, allowing them to leave charged tracks in the detector. For both scalar and fermionic scenarios, the produced charged components dominantly decay to DM (giving rise to missing energy) and charged pions, $\mathrm{DM}^\pm\to \mathrm{DM}^0\pi^\pm$~\cite{0706.4071}. This signal has been searched at the LHC that can exclude chargino masses as high as 1070 GeV~\cite{ATLAS:2022pib} with lifetimes $\tau\sim 10-100$ ns. However, as estimated in Ref.~\cite{hep-ph/0512090,0706.4071}, the typical lifetime of such charged particles that we are interested in is of order $\tau\sim 0.001-0.1$ ns, for which the above-mentioned LHC search is not applicable.

\vspace{0.5cm}
\textbf{3.\;Vectorlike Weak-singlet Charged Leptons:}--
Current collider constraints on vectorlike lepton iso-singlets are in the mass range between $125- 150\;\GeV$ at $95\%$ C.L.~\cite{2202.08676}, 
coming from direct searches with multiple leptons and/or b-tagged jets as final states. 
In addition, a characteristic signature of the vectorlike weak-singlet charged leptons is charged lepton flavour violation and non-universality induced by a modification of the $Z$- and $W$-couplings to leptons, which are however estimated to be outside the sensitivities of current experiments. The modification of the $Z$- and $W$-couplings comes from the mixing between the vectorlike states and the SM leptons. The same mixing also leads to quick decays of them, meaning that there are no obvious cosmological constraints on their properties.

\section{Conclusions}\label{sec:con}

Embedding inflation into GUTs is motivated by the potential proximity of the involved energy scales as well as by the necessity to dilute the monopoles produced from GUT symmetry breaking. Such an embedding usually employs an $R$-symmetry that enables a sufficiently flat field direction for slow-roll inflation. Examples include models of shifted or smooth versions of SUSY Hybrid Inflation, where the inflaton is a gauge singlet ``driving'' field, pseudosmooth realisations of Tribrid Inflation, where the inflaton resides in the (potentially gauge non-singlet) matter sector of the theory, and realisations of SUGRA New Inflation, hilltop inflation where the singlet component of adjoint 24-plet of $SU(5)$ plays the role of the inflaton while breaking the GUT symmetry. All such models can successfully realise inflation and do not produce monopoles or other topological defects when inflation ends (and efficiently dilute previously produced defects). 

It is known that when the $R$-symmetry is only broken by SUSY breaking effects in the inflation sector, there are comparatively light states in the particle spectrum. In the case of $SU(5)$ GUTs, these are a weak triplet and a colour octet from the adjoint representation, which acquire masses of the order of the SUSY scale (e.g.\ of some TeV). Although these light relics represent interesting signatures of the underlying GUT theory, potentially testable at colliders, they are often viewed as problematic, since they tend to disturb the otherwise ``automatic'' gauge coupling unification  in minimal SUSY extensions of the SM.

We have presented a novel ``Generalized Missing Partner Mechanism'' (GMPM), leading to a new scenario where this and other challenges of GUTs are resolved without excessive fine-tuning of parameters.
The GMPM splits the components of two extra vectorlike 10-plets, by generating their masses via ``missing partner’' couplings to a vectorlike 40-plet, resulting into light singly charged vectorlike component fields whereas the other 10-plet components are heavy. Together with light doublets from the ``Double Missing Partner Mechanism'' (DMPM), which realises doublet-triplet splitting of the 5-plets containing the MSSM Higgs representations, and the light colour octet and weak triplet states from the 24-plet, the $\beta$ functions of the gauge couplings change in a way that the ``automatic'' gauge coupling unification of the MSSM is maintained.  

Furthermore, the vectorlike 10-plets simultaneously enable realistic fermion mass relations, and the DMPM ensures that dimension five nucleon decay is suppressed. We also discussed present constraints on the scenario from particle physics experiments and cosmology. A discovery of signatures of the light relic states at future colliders or precision experiments could provide ``smoking gun'' signals of the scenario.

\appendix

\section{Explicit Decomposition of \texorpdfstring{$\overline{40}$}{40bar} into \texorpdfstring{$SU(5)$}{SU(5)} Tensors}\label{a:40b}
The representations $\overline{40}$ has three fundamental indices,
\begin{equation}
    \overline{40}^{[AB]C}\quad\text{with}\quad A,B,C=1,...,5\;,
\end{equation}
where the tensor is anti-symmetric under $A\leftrightarrow B$. We denote the individual fields as in \eqref{eq:40} and identify
\begin{align}
    \lambda_3&=\diag(1/2,-1/2,0,0,0),\nonumber\\
    \lambda_8&=\diag(1/(2\sqrt3),1/(2\sqrt3),-1/\sqrt3,0,0),\nonumber\\
    I_3&=\diag(0,0,0,1/2,-1/2),\label{eq:conv}\\
    Y&=\diag(-1/3,-1/3,-1/3,1/2,1/2)\nonumber.
\end{align}
Choosing a normalisation such that
\begin{equation}
    \overline{40}^{[AB]C}(\overline{40}^\dagger)_{[AB]C}=\overline Q^i\overline Q^\dagger_i+\overline U^{ci}\overline U^{c\dagger}_i+\Pi^i\Pi^\dagger_i+\Sigma^i\Sigma^\dagger_i+\Psi^i\Psi^\dagger_i+\Delta^i\Delta^\dagger_i,
\end{equation}
$\overline{40}^{[AB]C}$ is explicitly given by
\begin{align}
    \overline{40}^{[AB]1}&=\begin{pmatrix}
    0&\frac{\Sigma_1}{\sqrt2}&\frac{\Sigma_3}{\sqrt2}&\frac{\Psi_1}{\sqrt2}&\frac{\Psi_2}{\sqrt2}\\
    &0&\frac{\Sigma_7}{2\sqrt3}+\frac{\Sigma_8}{2}&\frac{\overline Q_1}{2\sqrt3}+\frac{\Psi_7}2&-\frac{\overline Q_3}{2\sqrt3}+\frac{\Psi_9}2\\
    &&0&-\frac{\overline Q_2}{2\sqrt3}+\frac{\Psi_8}2&\frac{\overline Q_5}{2\sqrt3}+\frac{\Psi_{11}}2\\
    &&&0&-\frac{\overline U^c_1}{\sqrt 3}\\
    &&&&0
    \end{pmatrix}\nonumber,\\
    \overline{40}^{[AB]2}&=\begin{pmatrix}
    0&\frac{\Sigma_2}{\sqrt2}&\frac{\Sigma_7}{\sqrt3}&-\frac{\overline Q_1}{2\sqrt3}+\frac{\Psi_7}2&\frac{\overline Q_3}{2\sqrt3}+\frac{\Psi_9}2\\
    &0&\frac{\Sigma_4}{\sqrt2}&\frac{\Psi_3}{\sqrt2}&\frac{\Psi_4}{\sqrt2}\\
    &&0&\frac{\overline Q_4}{2\sqrt3}+\frac{\Psi_{10}}2&-\frac{\overline Q_6}{2\sqrt3}+\frac{\Psi_{12}}2\\
    &&&0&-\frac{\overline U^c_2}{\sqrt 3}\\
    &&&&0
    \end{pmatrix}\nonumber,\\
    \overline{40}^{[AB]3}&=\begin{pmatrix}
     0&\frac{\Sigma_7}{2\sqrt3}-\frac{\Sigma_8}{2}&\frac{\Sigma_5}{\sqrt2}&\frac{\overline Q_2}{2\sqrt3}+\frac{\Psi_8}2&-\frac{\overline Q_5}{2\sqrt3}+\frac{\Psi_{11}}2\\
    &0&\frac{\Sigma_6}{\sqrt2}&-\frac{\overline Q_4}{2\sqrt3}+\frac{\Psi_{10}}2&\frac{\overline Q_6}{2\sqrt3}+\frac{\Psi_{12}}2\\
    &&0&\frac{\Psi_5}{\sqrt 2}&\frac{\Psi_6}{\sqrt 2}\\
    &&&0&-\frac{\overline U^c_3}{\sqrt 3}\\
    &&&&0
    \end{pmatrix}\label{eq:deco},\\
    \overline{40}^{[AB]4}&=\begin{pmatrix}
    0&-\frac{\overline Q_1}{\sqrt 3}&\frac{\overline Q_2}{\sqrt 3}&\frac{\Pi_1}{\sqrt 2}&-\frac{\overline U^c_1}{2\sqrt3}+\frac{\Pi_7}{2}\\
    &0&-\frac{\overline Q_4}{\sqrt 3}&\frac{\Pi_2}{\sqrt 2}&-\frac{\overline U^c_2}{2\sqrt3}+\frac{\Pi_8}{2}\\
    &&0&\frac{\Pi_4}{\sqrt 2}&-\frac{\overline U^c_3}{2\sqrt3}+\frac{\Pi_9}{2}\\
    &&&0&\frac{\Delta_1}{\sqrt 3}\\
    &&&&0
    \end{pmatrix}\nonumber,\\
    \overline{40}^{[AB]5}&=\begin{pmatrix}
    0&\frac{\overline Q_3}{\sqrt 3}&-\frac{\overline Q_5}{\sqrt 3}&\frac{\overline U^c_1}{2\sqrt3}+\frac{\Pi_7}{2}&\frac{\Pi_3}{\sqrt 2}\\
    &0&\frac{\overline Q_6}{\sqrt 3}&\frac{\overline U^c_2}{2\sqrt3}+\frac{\Pi_8}{2}&\frac{\Pi_5}{\sqrt 2}\\
    &&0&\frac{\overline U^c_3}{2\sqrt3}+\frac{\Pi_9}{2}&\frac{\Pi_6}{\sqrt 2}\\
    &&&0&\frac{\Delta_2}{\sqrt 2}\\
    &&&&0
    \end{pmatrix}.\nonumber
\end{align}
The individual quantum numbers under $\lambda_3,\,\lambda_8,\,I_3$ and $Y$ are given in Table \ref{tab:QN40}.

\begin{longtable}{|c|rrrr|}
        \caption{Charges of the individual components of $\overline{40}$ as given by eq. \eqref{eq:deco}. Our convention for the SM Casimirs are given in \eqref{eq:conv}.\label{tab:QN40}} \\ \hline
 \multicolumn{5}{| c |}{Begin of Table  \ref{tab:QN40}}\\\hline
 Fields&$\lambda_3$&$\lambda_8$&$I_3$&$Y$  \\\hline\hline
 \hline
 \endfirsthead

 \hline
 \multicolumn{5}{|c|}{Continuation of Table \ref{tab:QN40}}\\\hline
 Fields&$\lambda_3$&$\lambda_8$&$I_3$&$Y$  \\\hline\hline
 \hline
 \endhead

 \hline
 \endfoot

 \hline\hline
 \multicolumn{5}{| c |}{End of Table \ref{tab:QN40}}\\
 \hline
 \endlastfoot

         \multicolumn{5}{|l|}{$\overline{Q}(\overline 3,2)_{-1/6}$}\\\hline\hline
        $\overline Q_{1}$&$0$&$\frac{1}{\sqrt{3}}$&$\frac{1}{2}$&$-\frac{1}{6}$\\$\overline Q_{2}$&$\frac{1}{2}$&$-\frac{1}{2 \sqrt{3}}$&$\frac{1}{2}$&$-\frac{1}{6}$\\$\overline Q_{3}$&$0$&$\frac{1}{\sqrt{3}}$&$-\frac{1}{2}$&$-\frac{1}{6}$\\$\overline Q_{4}$&$-\frac{1}{2}$&$-\frac{1}{2 \sqrt{3}}$&$\frac{1}{2}$&$-\frac{1}{6}$\\$\overline Q_{5}$&$\frac{1}{2}$&$-\frac{1}{2 \sqrt{3}}$&$-\frac{1}{2}$&$-\frac{1}{6}$\\$\overline Q_{6}$&$-\frac{1}{2}$&$-\frac{1}{2 \sqrt{3}}$&$-\frac{1}{2}$&$-\frac{1}{6}$\\\hline\hline
         \multicolumn{5}{|l|}{$\overline U^c( 3,1)_{2/3}$}\\\hline\hline
         $\overline U^c_{1}$&$\frac{1}{2}$&$\frac{1}{2 \
        \sqrt{3}}$&$0$&$\frac{2}{3}$\\$\overline U^c_{2}$&$-\frac{1}{2}$&$\frac{1}{2 \
        \sqrt{3}}$&$0$&$\frac{2}{3}$\\$\overline U^c_{3}$&$0$&$-\frac{1}{\sqrt{3}}$&$0$&$\
        \frac{2}{3}$\\\hline\hline
         \multicolumn{5}{|l|}{$\Pi(\overline 3,3)_{2/3}$}\\\hline\hline
         $\Pi_1$&$\frac{1}{2}$&$\frac{1}{2 \sqrt{3}}$&$1$&$\frac{2}{3}$\\
         $\Pi_2$&$-\frac{1}{2}$&$\frac{1}{2 \sqrt{3}}$&$1$&$\frac{2}{3}$\\
         $\Pi_3$&$\frac{1}{2}$&$\frac{1}{2 \sqrt{3}}$&$-1$&$\frac{2}{3}$\\
        $\Pi_4$&$0$&$-\frac{1}{\sqrt{3}}$&$1$&$\frac{2}{3}$\\
         $\Pi_5$&$-\frac{1}{2}$&$\frac{1}{2 \sqrt{3}}$&$-1$&$\frac{2}{3}$\\
         $\Pi_6$&$0$&$-\frac{1}{\sqrt{3}}$&$-1$&$\frac{2}{3}$\\
         $\Pi_7$&$\frac{1}{2}$&$\frac{1}{2 \sqrt{3}}$&$0$&$\frac{2}{3}$\\
         $\Pi_8$&$-\frac{1}{2}$&$\frac{1}{2 \sqrt{3}}$&$0$&$\frac{2}{3}$\\
         $\Pi_9$&$0$&$-\frac{1}{\sqrt{3}}$&$0$&$\frac{2}{3}$\\\hline\hline
         \multicolumn{5}{|l|}{$\Sigma(8,1)_{-1}$}\\\hline\hline
        $\Sigma_1$&$\frac{1}{2}$&$\frac{\sqrt{3}}{2}$&$0$&$-1$\\$\Sigma_2$&$-\frac{1}{2}$&$\frac{\sqrt{3}}{2}$&$0$&$-1$\\$\Sigma_3$&$1$&$0$&$0$&$-1$\\$\Sigma_4$&$-1$&$0$&$0$&$-1$\\$\Sigma_5$&$\frac{1}{2}$&$-\frac{\sqrt{3}}{2}$&$0$&$-1$\\$\Sigma_6$&$-\frac{1}{2}$&$-\frac{\sqrt{3}}{2}$&$0$&$-1$\\$\Sigma_7$&$0$&$0$&$0$&$-1$\\$\Sigma_8$&$0$&$0$&$0$&$-1$\\\hline\hline
         \multicolumn{5}{|l|}{$\Psi(6,2)_{-1/6}$}\\\hline\hline
        $\Psi_{1}$&$1$&$\frac{1}{\sqrt{3}}$&$\frac{1}{2}$&$-\frac{1}{6}$\\$\Psi_{2}$&$1$&$\frac{1}{\sqrt{3}}$&$-\frac{1}{2}$&$-\frac{1}{6}$\\$\Psi_{3}$&$-1$&$\frac{1}{\sqrt{3}}$&$\frac{1}{2}$&$-\frac{1}{6}$\\$\Psi_{4}$&$-1$&$\frac{1}{\sqrt{3}}$&$-\frac{1}{2}$&$-\frac{1}{6}$\\$\Psi_{5}$&$0$&$-\frac{2}{\sqrt{3}}$&$\frac{1}{2}$&$-\frac{1}{6}$\\$\Psi_{6}$&$0$&$-\frac{2}{\sqrt{3}}$&$-\frac{1}{2}$&$-\frac{1}{6}$\\$\Psi_{7}$&$0$&$\frac{1}{\sqrt{3}}$&$\frac{1}{2}$&$-\frac{1}{6}$\\$\Psi_{8}$&$\frac{1}{2}$&$-\frac{1}{2 \sqrt{3}}$&$\frac{1}{2}$&$-\frac{1}{6}$\\$\Psi_{9}$&$0$&$\frac{1}{\sqrt{3}}$&$-\frac{1}{2}$&$-\frac{1}{6}$\\$\Psi_{10}$&$-\frac{1}{2}$&$-\frac{1}{2 \sqrt{3}}$&$\frac{1}{2}$&$-\frac{1}{6}$\\$\Psi_{11}$&$\frac{1}{2}$&$-\frac{1}{2 \sqrt{3}}$&$-\frac{1}{2}$&$-\frac{1}{6}$\\$\Psi_{12}$&$-\frac{1}{2}$&$-\frac{1}{2 \sqrt{3}}$&$-\frac{1}{2}$&$-\frac{1}{6}$\\\hline\hline
         \multicolumn{5}{|l|}{$\Delta(1,2)_{3/2}$}\\\hline\hline
         $\Delta_{1}$&$0$&$0$&$\frac{1}{2}$&$\frac{3}{2}$\\$\Delta_{2}$&$0$&$0$&$-\frac{1}{2}$&$\frac{3}{2}$\\\hline
    \end{longtable}
    We take the vev of $\Phi$ to be aligned along
    \begin{equation}
        \vev{\Phi}=v_{24}\diag (-2,-2,-2,3,3)/\sqrt{30}. \label{v24}
    \end{equation}
    Further we take the following decomposition for the Georgi-Glashow Model fields:
    \begin{align}
        10^{[AB]}&=\begin{pmatrix}
        0&u_3^c&-u_2^c&u_1&d_1\\
        &0&u_1^c&u_2&d_2\\
        &&0&u_3&d_3\\
        &&&0&e^c\\
        &&&&0
        \end{pmatrix},\\
        \overline 5_A&=\begin{pmatrix}d_1^c&d_2^c&d_3^c&e&-\nu \end{pmatrix}^T,\\
         (5_H)^A&=\begin{pmatrix}T_1&T_2&T_3&H^+_u&H^0_u \end{pmatrix}^T,\\
        (\overline 5_H)_A&=\begin{pmatrix}\overline T_1&\overline  T_2&\overline T_3&-H^-_d&H^0_d\end{pmatrix}^T.
    \end{align}

\section{Two-Loop \texorpdfstring{$\beta$}{beta}-Functions for Gauge Couplings}\label{a:gcu}
The two-loop RGE are given in Ref.~\cite{COLO-HEP-68} (cf. also~\cite{hep-ph/9311340} for an overview). For a direct product gauge group at two-loop we have
\begin{equation}
\mu\frac{\diff g_a}{\diff \mu}=\frac 1{(4\pi)^2}\beta_{g_a}^{(1)}+\frac 1{(4\pi)^4}\beta_{g_a}^{(2)},
\end{equation}
with
\begin{equation}
\beta_{g_a}^{(1)}=g_a^3\big(\sum_{R_a} S(R_a)-3C(G_a)\big),
\end{equation}
and 
\begin{equation}
\beta_{g_a}^{(2)}=g_a^3\Big\{-6g_a^2\big(C(G_a)\big)^2+2g_a^2C(G_a)\sum_{R_a}S(R_a)+4\sum_{R_a,R_b}g_b^2S(R_a)C(R_b)\Big\}+\mathcal O(g_a^3Y^2).
\end{equation}
Here, $S(R_a)$ denotes the Dynkin index summed over all chiral multiplets, whereas $C(R_a)$ ($C(G_a)$) is the quadratic Casimir invariant summed over all chiral (vector) multiplets.
Thus the two-loop gauge coupling RGEs for the SM gauge group can be written in terms of a three dimensional vector and a three by three matrix:
\begin{equation}
\mu\frac{\diff g_a}{\diff \mu}=\frac{1}{(4\pi)^2}b_ag_a^2+\frac{g_a^3}{(4\pi)^4}\sum_{b=1}^3B_{ab}g_b^2\;.
\end{equation}
The MSSM has the following values for $b_a$ and $B_{ab}$:
\begin{equation}
b_a^{MSSM}=\begin{pmatrix}33/5\\1\\-3\end{pmatrix}\quad\text{and}\quad B_{ab}^{MSSM}=\begin{pmatrix}\frac{199}{25}&\frac{27}5&\frac{88}5\\\frac95&25&24\\\frac{11}5&9&14\end{pmatrix} .\label{eq:rgefirst}
\end{equation}
The additional fields contribute in the following way:
\begin{align}
b_a^{\mathscr T}&=\begin{pmatrix}0\\2 \\0 \end{pmatrix}\quad\text{and}\quad B_{ab}^{\mathscr T}=\begin{pmatrix}0&0&0\\ 0&24&0\\0 &0&0\end{pmatrix}.
\\
b_a^{\mathscr O}&=\begin{pmatrix}0\\ 0\\ 3\end{pmatrix}\quad\text{and}\quad B_{ab}^{\mathscr O}=\begin{pmatrix}0&0&0\\ 0&0&0\\ 0&0&54\end{pmatrix}.
\\
b_a^{D'}&=\begin{pmatrix}3/10\\1/2 \\0 \end{pmatrix}\quad\text{and}\quad B_{ab}^{D'}=\begin{pmatrix}\frac9{50}&\frac9{10}&0\\\frac3{10} &\frac72&0\\0 &0&0\end{pmatrix}.
\\
b_a^{E^c_\alpha}&=\begin{pmatrix}3/5\\0 \\0 \end{pmatrix}\quad\text{and}\quad B_{ab}^{E^c_\alpha}=\begin{pmatrix}36/25&0&0\\ 0&0&0\\ 0&0&0\end{pmatrix}.
\\
b_a^{T_\alpha}&=\begin{pmatrix}1/5\\0 \\1/2 \end{pmatrix}\quad\text{and}\quad B_{ab}^{T_\alpha}=\begin{pmatrix}\frac4{75}&0&\frac{16}{15}\\ 0&0&0\\\frac2{15} &0&\frac{17}3\end{pmatrix}.
\\
b_a^{ U^c_\alpha\oplus\, Q_\alpha}&=\begin{pmatrix}9/10\\3/2 \\3/2 \end{pmatrix}\quad\text{and}\quad B_{ab}^{U^c_\alpha\oplus\, Q_\alpha}=\begin{pmatrix}\frac{43}{50}&\frac3{10}&\frac{24}{5}\\ \frac1{10}&\frac{21}2&8\\\frac35 &3&17\end{pmatrix}.\label{eq:rgelast}
\end{align}
Once the gauge couplings are run down to the SUSY scale they have to be converted from the $\overline {\mathrm{DR}}$ to the $\overline{\mathrm{MS}}$ normalisation scheme, which can be done using~\cite{hep-ph/9308222}
\begin{equation}
    g_{\overline{\mathrm{MS}}}=g_{\overline{\mathrm{DR}}}\bigg(1-\frac{g_{\overline{\mathrm{DR}}}^2}{96\pi^2}C(G)\bigg).
\end{equation}
\bibliographystyle{style}
\bibliography{REF,manual}

\end{document}